\documentclass[]{article}

\usepackage{hyperref}
\usepackage{amsmath}
\usepackage{amsfonts}
\usepackage{amsthm}
\usepackage{mathtools}
\usepackage{eucal} 
\usepackage{geometry} 
\usepackage{graphicx}
\usepackage{amssymb}
\usepackage{subcaption} 
\usepackage{authblk}
\usepackage[sorting=none]{biblatex}

\DeclareMathOperator*{\argmax}{arg\,max}
\newtheorem{theorem}{Theorem}

\begin{document}
\title{Parametric Constraints for Bayesian Knowledge Tracing from First Principles}

\author{Denis Shchepakin, Sreecharan Sankaranarayanan, Dawn Zimmaro}
\affil{Amazon, USA}

\maketitle

\begin{abstract}
Bayesian Knowledge Tracing (BKT) is a probabilistic model of a learner's state of mastery corresponding to a knowledge component. It considers the learner's state of mastery as a ``hidden" or latent binary variable and updates this state based on the observed correctness of the learner's response using parameters that represent transition probabilities between states. BKT is often represented as a Hidden Markov Model and the Expectation-Maximization (EM) algorithm is used to infer these parameters. However, this algorithm can suffer from several issues including producing multiple viable sets of parameters, settling into a local minima, producing degenerate parameter values, and a high computational cost during fitting. This paper takes a ``from first principles" approach to deriving constraints that can be imposed on the BKT parameter space. Starting from the basic mathematical truths of probability and building up to the behaviors expected of the BKT parameters in real systems, this paper presents a mathematical derivation that results in succinct constraints that can be imposed on the BKT parameter space. Since these constraints are necessary conditions, they can be applied prior to fitting in order to reduce computational cost and the likelihood of issues that can emerge from the EM procedure. In order to see that promise through, the paper further introduces a novel algorithm for estimating BKT parameters subject to the newly defined constraints. While the issue of degenerate parameter values has been reported previously, this paper is the first, to our best knowledge, to derive the constrains from first principles while also presenting an algorithm that respects those constraints.
\end{abstract}

\section{Introduction} 
Bayesian Knowledge Tracing (BKT) \cite{corbett1994knowledge} was introduced as a way to model the changing knowledge states of students who were interacting with an adaptive learning system for skill acquisition. To this day, it remains the most widespread way to model student learning predominantly owing to its sufficient complexity for many use cases \cite{craig2013impact, minn2022ai, kabudi2021ai, liu2022knowledge}. The model considers the learner's state of mastery as a ``hidden" or latent binary variable with two possible states {--} Mastery and Non-Mastery (Sometimes called Proficient and Not-yet-proficient \cite{bhatt2020evaluating, kim2023variational}, or Knows and Does-not-know \cite{hawkins2014learning}). It then uses four parameters {--} the initial probability of mastery, the transition probability from non-mastery to mastery over a single learning opportunity, the probability of a correct answer with the learner in the non-mastery state (guess), and the probability of an incorrect answer with the learner in the mastery state (slip), to ``predict" whether a given learner is in the mastery state or not. In order to learn the value of these parameters, BKT is most often represented as a Hidden Markov Model \cite{beck2007identifiability}, and the parameters are determined using an Expectation-Maximization (EM) algorithm \cite{dempster1977maximum, chang2006bayes} on historical data.

As with any other EM algorithm, this may result in multiple sets of (highly dissimilar) parameters that fit the data equally well \cite{beck2007identifiability} which, in the case of BKT, affects interpretability. Further, the parameters may be degenerate, i.e., fit the data but violate certain assumptions leading to incorrect decisions if used in real systems \cite{baker2008more}. Finally, if the algorithm needs to be re-run after it is post-hoc determined as producing degenerate parameters, then that becomes computationally expensive. Several approaches have been suggested which can partially resolve the issue including determining the starting values that lead to degenerate parameters \cite{pardos2010navigating} (and avoiding them), computing Dirichlet priors for each parameter and using that to bias the search \cite{wang2013class}, clustering parameters across similar skills \cite{ritter2009reducing}, and machine-learned models for some of the parameters \cite{baker2008more}. Approaches that provably avoid degenerate parameters have also been discussed in literature but they instead sacrifice the precision provided by the EM algorithm \cite{hawkins2014learning}. Thus, in this paper, we first derive parametric constraints for the BKT parameters from first principles, that, if satisfied, necessarily avoid degenerate parameters. Then, we present a novel EM algorithm that respects the derived constraints thus allowing them to be used in practice. 

While similar constraints have previously emerged by studying fixed points of the BKT model \cite{van2013properties}, here we derive them from first principles applied to the logic of the modeled process. Moreover, we prove that these less strict constrains are sufficient compared to the ones derived in \cite{van2013properties}. Finally, we also present a novel EM algorithm that respects these constraints.

\section{Defining the BKT Model}\label{section:BKT:definition}
BKT assumes that for each knowledge component (KC), a learner can be in either the Proficient or Not-yet-proficient state at a given point in time. After attempting an assessment, the learner receives feedback, either explicitly or gleans it from the fact that their response was marked correct or incorrect. This is an opportunity to become proficient in the corresponding knowledge component. If the learner learns successfully, they transition from the not-yet-proficient state to the proficient state for that corresponding KC. Once the learner becomes proficient in a KC, they cannot transition back to the not-yet-proficient state (Note that variations of the BKT model that incorporate ``forgetting" allow this transition \cite{khajah2016deep}). A BKT model is then constructed and applied for each KC independently.

Let $L^{(d)}_t$ be an event that learner $d$ is proficient after receiving $t$ rounds of feedback; $C^{(d)}_t$ is an event that learner $d$ answers assessment $t$ correctly; $G^{(d)}_t$ is an event that a learner $d$ guesses a correct answer for an assessment $t$ while not being proficient; $S^{(d)}_t$ is an event that a learner $d$ makes a mistake (``slips") at an assessment $t$ while being proficient; and $R^{(d)}_t$ is an event that a non-proficient learner $d$ transitions to a proficient state after receiving $t$ rounds of feedback. The classic BKT model assumes that probabilities of guess, slip, and transition events are independent of the learner and the assessment, and depend only on the learner being proficient. Moreover, the initial proficiency probability $P(L^{(d)}_0)$ is also assumed to be independent from the learner, and, rather, a proportion of learners in the population that are proficient before attempting their first assessment is used. So, we will omit redundant upper and lower indexes for $G$, $S$, $R$, and $L_0$ events and their probabilities. See Figure \ref{fig:bkt} for an outline of the model.
\begin{figure}
    \centering
	\begin{tabular}{|ccl|}
		\hline
		&& \\
		$P(L_0)$ & Prior Proficiency & the probability the learner is in proficient \\ 
		& & state for the KC prior to first feedback.  \\
		&&\\
		$P(G)$ & Guess & the probability of correct guess at\\
		& & an assessment while not being proficient \\
		& & at the corresponding KC. \\
		&&\\
		
		$P(S)$ & Slip & the probability of slip (mistake) at \\
		& &  an assessment while being proficient at\\
		& &  the corresponding KC. \\
		&&\\
		
		$P(R)$ & Transition & the probability of a learner becoming\\ 
		& & proficient in KC after making an attempt\\
		& &   at an assessment and reading the feedback. \\
		&&\\
		\hline
		&&\\
		
		$P(L^{(d)}_t)$ & Proficiency & the probability of learner $d$ being in a proficient\\
		& & state after receiving $t$ rounds of feedback.\\
		&& \\
		$P(C^{(d)}_t)$ & Correctness of & the probability of learner $d$ answering\\
		& an Attempt &  assessment $t$ correctly.\\
		
		&&\\
		\hline
	\end{tabular}
	\caption{BKT parameters and notation}\label{fig:bkt}
\end{figure}

The BKT model defines $P(C^{(d)}_{t+1})$ thus {--}
\begin{equation}\label{eq:bkt:correct_answer}
	P(C^{(d)}_{t+1}) = P(L^{(d)}_t) \cdot (1 - P(S)) + ( 1- P(L^{(d)}_t) ) \cdot P(G)
\end{equation}
Using the Bayes' rule, we get {--}
\begin{equation}\label{eq:bkt:bayes_update}
	\begin{array}{c}
		P(L^{(d)}_t|C^{(d)}_{t+1}) = \dfrac{P(L^{(d)}_t) \cdot ( 1 - P(S) ) }{ P(L^{(d)}_t) \cdot ( 1 - P(S) ) + ( 1 - P(L^{(d)}_t) ) \cdot P(G)}, \\
		\\
		P(L^{(d)}_t|\overline{C^{(d)}_{t+1}}) = \dfrac{P(L^{(d)}_t) \cdot P(S)}{P(L^{(d)}_t) \cdot P(S) + ( 1 - P(L^{(d)}_t) ) \cdot ( 1 - P(G) )}
	\end{array}
\end{equation}
where $\overline{C^{(d)}_t}$ is an event complementary to $C^{(d)}_t$, i.e., an event that learner $d$ answers assessment $t$ incorrectly. After an attempt at the assessment and receiving a feedback, the learner has a chance of transitioning if they are not already proficient {--}
\begin{equation}\label{eq:bkt:transitioning}
	P(L^{(d)}_{t+1}) = P(L^{(d)}_t | \boldsymbol{\cdot} ) + P(R) \cdot ( 1 - P(L^{(d)}_t | \boldsymbol{\cdot} ) )
\end{equation}
where $P(L^{(d)}_t | \boldsymbol{\cdot} )$ is either $P(L^{(d)}_t|C^{(d)}_t)$ or $P(L^{(d)}_t|\overline{C^{(d)}_t})$ depending on the collected data for the learner.

Knowing the values of all parameters of the BKT model will allow us to predict the probability of learner $d$ being proficient in the KC, $P(L^{(d)}_t)$ (which we will refer to simply as ``proficiency" of learner $d$).
\section{Restrictions on the BKT Parameters}
Prior to estimating the BKT parameters, we need to place some restrictions on them to maintain the logic of the modeled process when used in real systems. All results obtained in this section are not learner specific. Thus, for the sake of readability, we will omit all learner-specific indexes in this section, e.g., we will use $L_t$ instead of $L^{(d)}_t$. 

First, it does not make sense for $P(S)$ and $P(G)$ to be $0$ since that would simply eliminate their use as parameters entirely. $P(G)$ being $1$ would mean that a learner in the non-mastery state would necessarily guess and get the question right each time which is unrealistic. Similarity, $P(S)$ being equal to $1$ would mean that a learner in the mastery state would necessarily slip and get the question wrong each time which obviates the very definition of mastery. Thus, our first constraint is that $P(S)$ and $P(G)$ both vary between $0$ and $1$ without ever taking the extreme values exactly. Next, $P(R)$ is also between $0$ and $1$. If $P(R) = 0$, then learners cannot transition, and the learning experience is a priori useless. if $P(R) = 1$, then the learning experience has a 100\% success rate. Both situations cannot be guaranteed. Next, if $P(L_0) = 0$, then from (\ref{eq:bkt:bayes_update}) - (\ref{eq:bkt:transitioning}), all $P(L_t) = 0$, which is  an uninteresting scenario to consider. Moreover, if $P(L_t) = 0$, then from (\ref{eq:bkt:transitioning}), it follows that $P(L_{t - 1} | \boldsymbol{\cdot}) = 0$. And from (\ref{eq:bkt:bayes_update}), we can see it is possible only if $P(L_{t-1}) = 0$. Therefore, by induction, $P(L_0) = 0$.  Similarly, $P(L_t) = 1$ if and only if $P(L_0) = 1$. Thus, we can assume {--}
\begin{enumerate}
	\item $0 < P(G) < 1$,
	\item $0 < P(S) < 1$,
	\item $0 < P(R) < 1$,
	\item $0 < P(L_t) < 1$ for all $t=0,\cdots,T$.
\end{enumerate}
There are some additional restrictions we can add for the BKT parameters. Namely, we want correct answers to increase our estimate of learner's proficiency both before and after the transition. Similarly, incorrect answers should lower the proficiency.
\begin{equation}\label{eq:bkt:condition_pre-transition_general}
	P(L_t | C_{t+1}) \geq P(L_t) \geq P(L_t | \overline{C_{t+1}})
\end{equation}
and
\begin{equation}\label{eq:bkt:condition_post-transition_general}
	P(L_{t + 1} | C_{t+1}) \geq P(L_t) \geq P(L_{t + 1} | \overline{C_{t+1}})
\end{equation}

The inequalities in (\ref{eq:bkt:condition_pre-transition_general}) yield a natural restriction on the parameters {--}
\begin{equation}
	1 - P(S) \geq P(G)
\end{equation}
that is, the probability of answering an assessment correctly is higher if a learner is proficient. Moreover, this restriction is also sufficient for the left inequality in (\ref{eq:bkt:condition_post-transition_general}) to be true. Let us prove these statements.
\begin{proof}
	Let us consider the first three inequalities.
	\begin{enumerate}
		\item $P(L_t|C_{t+1}) \geq P(L_t)$:
		$$\dfrac{P(L_t) \cdot ( 1 - P(S) ) }{ P(L_t) \cdot ( 1 - P(S) ) + P(G) \cdot ( 1 - P(L_t) )} \geq P(L_t),$$
		$$\dfrac{ 1 - P(S) }{ P(L_t) \cdot ( 1 - P(S) ) + P(G) \cdot ( 1 - P(L_t) )} \geq 1,$$
		$$1 - P(S) \geq P(L_t) \cdot ( 1 - P(S) ) + P(G) \cdot ( 1 - P(L_t) ),$$
		$$1 - P(S) - P(G) \geq P(L_t) \cdot ( 1 - P(S) -P(G)),$$
		$$(1 - P(S) - P(G)) \cdot (1 - P(L_t)) \geq 0,$$
		which is always true if and only if $1 - P(S) - P(G) \geq 0$.
		
		\item $P(L_t) \geq P(L_t|\overline{C_{t+1})}$:
		$$P(L_t) \geq \dfrac{P(L_t) \cdot P(S)}{P(L_t) \cdot P(S) + ( 1 - P(L_t) ) \cdot ( 1 - P(G) )},$$
		$$1 \geq \dfrac{P(S)}{P(L_t) \cdot P(S) + ( 1 - P(L_t) ) \cdot ( 1 - P(G) )},$$
		$$P(L_t) \cdot P(S) + ( 1 - P(L_t) ) \cdot ( 1 - P(G) ) \geq P(S),$$
		$$(1 - P(L_t)) \cdot (1 - P(G) - P(S)) \geq 0,$$
		which is always true if and only if $1 - P(S) - P(G) \geq 0$.
		
		\item $P(L_{t + 1} | C_{t+1}) \geq P(L_t)$:
		$$P(L_{t+1} | C_{t+1}) = P(L_t | C_{t+1} ) + P(R) \cdot ( 1 - P(L_t | C_{t+1} ) ) \geq P(L_t),$$
		$$P(L_t | C_{t+1} ) \cdot (1 - P(R)) + P(R) \geq P(L_t),$$
		$$\dfrac{P(L_t) \cdot ( 1 - P(S) ) \cdot (1 - P(R))}{ P(L_t) \cdot ( 1 - P(S) ) + P(G) \cdot ( 1 - P(L_t) )} + P(R) \geq P(L_t),$$
		$$\dfrac{P(L_t) \cdot ( 1 - P(S) ) + P(G) \cdot ( 1 - P(L_t) ) \cdot P(R)}{ P(L_t) \cdot ( 1 - P(S) ) + P(G) \cdot ( 1 - P(L_t) )} \geq P(L_t),$$
		\begin{multline*}
			P(L_t) \cdot ( 1 - P(S) ) + P(G) \cdot ( 1 - P(L_t) ) \cdot P(R) \geq\\
			P(L_t)^2 \cdot ( 1 - P(S) ) + P(G) \cdot P(L_t) \cdot ( 1 - P(L_t) ),
		\end{multline*}
		$$P(L_t) \cdot (1 - P(S)) \cdot (1 - P(L_t)) + (1 - P(L_t)) \cdot P(G) \cdot (P(R) - P(L_t)) \geq 0,$$
		$$P(L_t) \cdot (1 - P(S)) + P(G) \cdot (P(R) - P(L_t)) \geq 0,$$
		$$P(L_t) \cdot (1 - P(S) - P(G)) + P(G) \cdot P(R) \geq 0,$$
		which is true if $1 - P(S) - P(G) \geq 0$.		 
	\end{enumerate}
\end{proof}

The final inequality in (\ref{eq:bkt:condition_post-transition_general}) yields a non-trivial restriction {--}
\begin{equation}\label{eq:bkt:steady_state_general}
	P(L_t) \geq \dfrac{( 1 - P(G) ) \cdot P(R)}{1 - P(S) - P(G)}.
\end{equation}
\begin{proof}
	$$P(L_t) \geq P(L_{t + 1} | \overline{C_{t+1}}) = P(L_t | \overline{C_{t+1}} ) + P(R) \cdot ( 1 - P(L_t | \overline{C_{t+1}} ) ),$$
	$$P(L_t) \geq P(L_t | \overline{C_{t+1}} ) \cdot (1 - P(R)) + P(R),$$
	$$P(L_t) \geq \dfrac{P(L_t) \cdot P(S) \cdot (1 - P(R))}{P(L_t) \cdot P(S) + ( 1 - P(L_t) ) \cdot ( 1 - P(G) )} + P(R),$$
	$$P(L_t) \geq \dfrac{P(L_t) \cdot P(S) + ( 1 - P(L_t) ) \cdot ( 1 - P(G) ) \cdot P(R)}{P(L_t) \cdot P(S) + ( 1 - P(L_t) ) \cdot ( 1 - P(G) )},$$
	\begin{multline*}
		P(L_t)^2 \cdot P(S) + P(L_t) \cdot ( 1 - P(L_t) ) \cdot ( 1 - P(G) ) \geq\\
		P(L_t) \cdot P(S) + ( 1 - P(L_t) ) \cdot ( 1 - P(G) ) \cdot P(R),
	\end{multline*}
	$$P(L_t) \cdot P(S) \cdot (1 - P(L_t)) + (1 - P(L_t)) \cdot (1 - P(G)) \cdot (P(R) - P(L_t)) \leq 0,$$
	$$P(L_t) \cdot P(S) + (1 - P(G)) \cdot (P(R) - P(L_t)) \leq 0,$$
	$$P(L_t) \cdot (1 - P(G) - P(S)) \geq (1 - P(G)) \cdot P(R).$$
	Note that $1 - P(G) - P(S) \neq 0$, otherwise $P(G) = 1$ or $P(R) = 0$. Therefore,
	$$P(L_t) \geq \dfrac{( 1 - P(G) ) \cdot P(R)}{1 - P(S) - P(G)}.$$
\end{proof}

Let us define the value in the right-hand side of (\ref{eq:bkt:steady_state_general}) as $P^*$. It can be shown that if $P^* < P(L_{t^*}) < 1$, then $P^* < P(L_t) < 1$ for any $t > t^*$ and any sequence of attempts. Namely, the following is true {--}
\begin{theorem}\label{th:sequence}
In a sequence of all failed attempts $P(L_t)$ will asymptotically approach $P^*$ from the right, and in a sequence of all successful attempts $P(L_t)$ will asymptotically approach $1$ from the left.
\end{theorem}
\begin{proof}
	Let us consider a sequence of only failed attempts $F = (F_1, F_2, \cdots , F_T)$. And let $P(F_t) = P^* + \varepsilon / (1 - P(S) - P(G))$ for some $0 < \varepsilon < (1 - P(S) - P(G)) \cdot (1 - P^*)$. Then,
	\begin{multline}\label{eq:bkt:proof_ss_low}
		P(F_{t + 1}) = \dfrac{P(F_t)\cdot P(S)}{P(F_t) \cdot P(S) + (1 - P(F_t)) \cdot (1 - P(G))} (1 - P(R)) + P(R)\\\\
		= \dfrac{\dfrac{(1 - P(G)) \cdot P(R) + \varepsilon}{1 - P(S) - P(G)} \cdot P(S) \cdot (1 - P(R))}{\dfrac{(1 - P(G)) \cdot P(R) + \varepsilon}{1 - P(S) - P(G)} \cdot P(S) + \left( 1 - \dfrac{(1 - P(G)) \cdot P(R) + \varepsilon}{1 - P(S) - P(G)} \right) \cdot (1 - P(G))}\\\\+ P(R)\\\\
		= \dfrac{((1 - P(G)) \cdot P(R) + \varepsilon) \cdot P(S) \cdot (1 - P(R))}{\splitfrac{((1 - P(G)) \cdot P(R) + \varepsilon) \cdot P(S)}{+ (1 - P(S) - P(G) - P(R) + P(G) \cdot P(R) - \varepsilon) \cdot (1 - P(G))}} + P(R)\\\\
		= \dfrac{P(R) \cdot P(S) \cdot \left[ (1 - P(G)) \cdot (1 - P(R)) - \varepsilon \right] + \varepsilon \cdot P(S)}{\splitfrac{(1 - P(G)) \cdot P(R) \cdot (P(S) - 1 + P(G))}{+ (1 - P(G)) \cdot (1 - P(S) - P(G)) + \varepsilon \cdot (P(S) - 1 + P(G))}} + P(R)\\\\
		= \dfrac{P(R) \cdot P(S) \cdot \left[ (1 - P(G)) \cdot (1 - P(R)) - \varepsilon \right] + \varepsilon \cdot P(S)}{(1 - P(S) - P(G)) \cdot \left[ (1 - P(G)) \cdot (1 - P(R)) - \varepsilon \right]} +P(R)\\\\
		= \dfrac{P(R) \cdot P(S)}{1 - P(S) - P(G)} + \dfrac{\varepsilon \cdot P(S)}{(1 - P(S) - P(G)) \cdot \left[ (1 - P(G)) \cdot (1 - P(R)) - \varepsilon \right]} + P(R)\\\\
		= P^* + \dfrac{\varepsilon \cdot P(S)}{(1 - P(S) - P(G)) \cdot \left[ (1 - P(G)) \cdot (1 - P(R)) - \varepsilon \right]}.
	\end{multline}
	Note that $P(F_t) < 1$, so by the definition of $\varepsilon$, we have {--}
	$$\left[(1 - P(G)) \cdot (1 - P(R)) - \varepsilon \right]> P(S) > 0.$$
	Thus, from (\ref{eq:bkt:proof_ss_low})
	$$P^* < P(F_{t+1}) < P^* + \dfrac{\varepsilon}{1 - P(S) - P(G)} = P(F_t).$$
	Therefore, if $P(F_0) > P^*$, then the sequence $(P(F_1), P(F_1), \cdots , P(F_T))$ asymptotically approaches $P^*$ from the right.
	
	Let us now consider a sequence of only successful attempts $U = (U_1, U_2, \cdots, U_T)$. We know that the sequence of proficiencies $(P(U_1), P(U_2), \cdots , P(U_T))$ is increasing and cannot reach $1$ from previous discussion. Let us show, that it asymptotically approaches $1$.  Let $P(U_t) = 1 - \epsilon$ for some $0 < \epsilon < 1$. Then,
	
	\begin{multline}\label{eq:bkt:proof_ss_high}
		P(U_{t+1}) = \dfrac{P(U_t) \cdot (1 - P(S)) }{P(U_t) \cdot (1 - P(S)) + (1 - P(U_t)) \cdot P(G)} (1 - P(R)) + P(R)\\\\		
		= \dfrac{P(U_t) \cdot (1 - P(S)) + (1 - P(U_t)) \cdot P(G) \cdot P(R)}{P(U_t) \cdot (1 - P(S)) + (1 - P(U_t)) \cdot P(G)} =\\\\
		= \dfrac{(1 - \epsilon) \cdot (1 - P(S)) + \epsilon \cdot P(G) \cdot P(R)}{(1 - \epsilon) \cdot (1 - P(S)) + \epsilon \cdot P(G))} = 1 - \dfrac{\epsilon \cdot P(G) \cdot (1 - P(R))}{(1 - \epsilon) \cdot (1 - P(S)) + \epsilon \cdot P(G))}.
	\end{multline}
	Note that {--}
	$$(1 - \epsilon) \cdot (1 - P(S) - P(G)) + P(R) \cdot P(G) > 0,$$
	$$(1 - \epsilon) \cdot (1 - P(S)) - P(G) + \epsilon \cdot P(G) + P(R) \cdot P(G) > 0,$$
	$$(1 - \epsilon) \cdot (1 - P(S)) + \epsilon \cdot P(G) > P(G) \cdot (1 - P(R)),$$
	$$\dfrac{P(G) \cdot (1 - P(R))}{(1 - \epsilon) \cdot (1 - P(S)) + \epsilon \cdot P(G)} < 1.$$
	Using (\ref{eq:bkt:proof_ss_high}), gives us
	$$P(U_t) = 1 - \epsilon < P(U_{t+1}) < 1.$$
	Therefore, if $P(U_0) < 1$, then the sequence $(P(U_1), P(U_2), \cdots, P(U_T))$ asymptotically approaches $1$ from the left.
	
	Finally, for any sequence of attempts $L = (L_1, L_2, \cdots, L_T)$ if $P(L_0) = P(F_0) = P(U_0)$, then sequence $F$ is the lower bound for $L$, and sequence $U$ is the upper bound for $L$.
\end{proof}

Thus, we arrive at the following succinct list of restrictions on the BKT model parameters {--}
\begin{equation}\label{eq:bkt:final_cond_G}
	0 < P(G) < 1,
\end{equation}

\begin{equation}\label{eq:bkt:final_cond_S}
	0 < P(S) < 1,
\end{equation}

\begin{equation}\label{eq:bkt:final_cond_R}
	0 < P(R) < 1,
\end{equation}

\begin{equation}\label{eq:bkt:final_cond_SG}
	1 - P(S) - P(G) \geq 0,
\end{equation}

\begin{equation}\label{eq:bkt:final_cond_ss}
	\dfrac{( 1 - P(G) ) \cdot P(R)}{1 - P(S) - P(G)} < P(L_0) < 1.
\end{equation}

\section{Estimating the Parameters}
Let $X = (X^{(1)}, X^{(2)}, \cdots , X^{(D)})$ be a hidden process of learners' states of proficiency where $X^{(d)} = (X^{(d)}_1, X^{(d)}_2, \cdots, X^{(d)}_{T^{(d)}})$ is a sequence corresponding to learner $d$. $X^{(d)}_t$ takes value $0$ if learner $d$ is not proficient after $t$ rounds of feedback, and value $1$ if learner $d$ is proficient after $t$ rounds of feedback. Analogously, let $Y= ( Y^{(1)}, Y^{(2)}, \cdots , Y^{(D)} )$ be an observable process of a learner's attempts at assessments, with $Y^{(d)}= ( Y_1^{(d)} , Y_2^{(d)} , \cdots , Y^{(d)}_{T^{(d)}})$ is a sequence of attempts corresponding to a learner $d$. Similarly, $Y^{(d)}_t$ takes value $0$ and $1$ for incorrect and correct answers by learner $d$ at assessment $t$, respectively. And let $y$ be a realization of $Y$, i.e., an available dataset. We cannot directly observe a corresponding realization of the latent process $X$. For maximum likelihood estimation of parameters we would need to maximize the marginal likelihood function with respect to $X$, which is intractable. Let us consider approaches to estimate the parameters of the BKT model for a given KC: $P(L_0)$, $P(G)$, $P(S)$, and $P(R)$.

\subsection{Expectation-Maximization Algorithm}
Expectation–maximization (EM) algorithm \cite{dempster1977maximum} is an iterative algorithm to find local maximum likelihood estimates of parameters in a model with unobserved latent variables. EM is not guaranteed to converge to an optimal solution, but rather to a local optimal solution. EM is used when computation of the likelihood function is intractable due to presence of latent variables. Starting with a random initial guess for parameters, the algorithm improves the estimate for model parameters in each iteration by guaranteeing that the new parameter values correspond to higher values of the log-likelihood function without explicitly computing it. Let us denote $\theta$ as a vector of all model parameters, and $\theta^*$ as the parameter estimates at the current step of EM. During each step, the function $Q(\theta) = Q(\theta | \theta^*)$ is constructed as the expected value of log-likelihood function of $\theta$ with respect to the current conditional distribution of $X$ given data $y$ and $\theta^*$ {--}
\begin{equation}\label{eq:EM:Q_general}
	Q(\theta | \theta^*) = \mathbb{E}_{X|y, \theta^*} \left[ \log P(y, X | \theta) \right].
\end{equation}
EM states that {--}
\begin{equation}\label{eq:EM:main_property}
	\forall \theta : \log P (y | \theta) -  \log P (y | \theta^*) \geq Q(\theta | \theta^*) - Q(\theta^* | \theta^*),
\end{equation}
that is, any $\theta$ that increases value of $Q$ over $Q(\theta^* | \theta^*)$ also improves the value of the corresponding marginal log-likelihood function. EM defines the next value of $\theta^*$ as the {--}
\begin{equation}\label{eq:EM:argmax_Q}
	\theta^*_{\text{next}} = \argmax\limits_\theta Q(\theta | \theta^*).
\end{equation}

For a discrete case, like a BKT model, (\ref{eq:EM:Q_general}) becomes {--}
\begin{equation}\label{eq:EM:Q_discrete}
	Q(\theta | \theta^*) = \sum\limits_{x \in \mathcal{X}} \log \left[ P(y, x | \theta )\right] \cdot P(x | y, \theta^*),
\end{equation}
where $\mathcal{X}$ is a set of all possible $X$. Because $P(y, x | \theta^*) = P(x | y, \theta^*) \cdot P(y | \theta^*)$ and $P(y | \theta^*)$ is a constant with respect to $\theta$, maximization of (\ref{eq:EM:Q_discrete}) is equivalent to a maximization of {--}
\begin{equation}\label{eq:EM:Q_hat_discrete}
	\widehat{Q}(\theta | \theta^*) = \sum\limits_{x \in \mathcal{X}} \log \left[ P(y, x | \theta )\right] \cdot P(y, x | \theta^*).
\end{equation}
Sometimes, maximization of (\ref{eq:EM:Q_hat_discrete}) is more convenient than (\ref{eq:EM:Q_discrete}).

BKT can be modeled as a Hidden Markov model and, therefore, a special case of EM algorithm, Baum-Welch algorithm \cite{baum1970maximization}, can be used. The Baum-Welch algorithm provides closed forms for $\theta^*_{\text{next}}$ based on $\theta^*$ values. It is fully described in Appendix \ref{apdx:Baum}.

Note, that the Baum-Welch algorithm does not guarantee that (\ref{eq:bkt:final_cond_SG}) - (\ref{eq:bkt:final_cond_ss}) are satisfied. To avoid cases where Baum-Welch converges to a unsuitable parameters (i.e., degenerate parameters), we offer to use a different approach.

\subsubsection{Novel EM Algorithm using the Interior-Point Method}
We want an algorithm that will always yield meaningful parameters for the BKT model by satisfying conditions (\ref{eq:bkt:final_cond_G}) - (\ref{eq:bkt:final_cond_ss}). That can be achieved if instead of just maximizing $\widehat{Q}$ in (\ref{eq:EM:argmax_Q}), we maximize it under (\ref{eq:bkt:final_cond_G}) - (\ref{eq:bkt:final_cond_ss}) restrictions. Note that the corresponding log-likelihood function will increase due to property (\ref{eq:EM:main_property}). Conditions (\ref{eq:bkt:final_cond_G}) - (\ref{eq:bkt:final_cond_R}) and right-hand side of (\ref{eq:bkt:final_cond_ss}) are satisfied automatically due to the form of $\log$ functions in $\widehat{Q}$, see (\ref{eq:BKT:Q}). Finally, (\ref{eq:bkt:final_cond_SG}) and left-hand side of (\ref{eq:bkt:final_cond_ss}) can be combined into a single inequality, resulting in the following non-linear optimization problem {--}
\begin{equation}\label{eq:EM_interior:KKT_problem}
	\begin{array}{lcl}
		\theta^*_{\text{next}} & = & \argmax\limits_\theta \widehat{Q}(\theta | \theta^*),\\\\
		& \text{s.t.} & (1 - P(S) - P(G))  \cdot P(L_0) - (1 - P(G)) \cdot P(R) \geq 0,
	\end{array}
\end{equation}
where $ \widehat{Q}$ has form (\ref{eq:BKT:Q}). We will use the interior-point method on  (\ref{eq:EM_interior:KKT_problem}). The goal is to find the maximum of the barrier function {--}
\begin{equation}\label{eq:EM_interior:B}
	B(\theta, \mu) = \widehat{Q}(\theta | \theta^*) + \mu \cdot \log c(\theta),
\end{equation}
where $c(\theta)$ is the left-hand side of the constrain from (\ref{eq:EM_interior:KKT_problem}), and $\mu$ is a so-called barrier parameter. We will iterate through a decreasing sequence of values for $\mu$ parameter $\mu_1 > \mu_2 > \cdots > \mu_W = 0$, finding a maximum of $W$ in each iteration. As $\mu$ approaches $0$, the maximum of $B$ converges to the solution of (\ref{eq:EM_interior:KKT_problem}). Next, a dual variable $\lambda$ is introduced, defined as $c(\theta) \cdot \lambda = \mu$. To find the extremum point of $B$ we need to find zero of the following vector function {--}
\begin{equation}
	F = \left[
	\begin{array}{c}
		\nabla B\\
		\lambda \cdot c(\theta) - \mu
	\end{array}\right].
\end{equation}
We will use a Newton's method to find zero of $F$. We start with some initial guesses $\theta_1$ and $\lambda_1$ by solving and update them by solving {--}
\begin{equation}\label{eq:EM_interior:Newton_method}
	J_F(\theta_k, \lambda_k) \times \left[
	\begin{array}{c}
		\Delta \theta\\
		\Delta \lambda
	\end{array}\right] = - F(\theta_k, \lambda_k),
\end{equation}
where $J_F$ is a Jacobian of $F$; and updating {--}
\begin{equation}\label{eq:EM_interior:Newton_update}
	\begin{array}{clc}
		\theta_{k+1} = \theta_k + \nu \cdot \Delta \theta,\\
		\lambda_{k+1} = \lambda_k + \nu \cdot \Delta \lambda,
	\end{array}
\end{equation}
where $\nu$ is a value small enough, so updated $\theta_{k+1}$ and $\lambda_{k+1}$ satisfy $c(\theta_{k+1}) \geq 0$ and $\lambda_{k+1} \geq 0$. Next,
\begin{equation}\label{eq:EM_interior:F}
	F = \left[
	\begin{array}{ccl}
		\dfrac{\partial \widehat{Q}}{\partial P(L_0)} &+& \lambda \cdot (1 - P(S) - P(G))\\\\
		\dfrac{\partial \widehat{Q}}{\partial P(G)} &+&  \lambda \cdot (P(R) - P(L_0))\\\\
		\dfrac{\partial \widehat{Q}}{\partial P(S)} &-& \lambda \cdot P(L_0)\\\\
		\dfrac{\partial \widehat{Q}}{\partial P(R)} &-& \lambda \cdot (1 - P(G))\\\\
		\multicolumn{3}{l}{\lambda \cdot (1 - P(S) - P(G)) \cdot P(L_0) - \lambda \cdot (1 - P(G)) \cdot P(R) - \mu}
	\end{array}\right],
\end{equation}
where partial derivatives of $\widehat{Q}$ are given by (\ref{eq:BKT:dQdL_0}) - (\ref{eq:BKT:dQdR}). And {--}
\begin{equation}\label{eq:EM_interior:JF}
	J_F = \scalebox{0.9}{$\left[
	\begin{array}{ccccc}
		\dfrac{\partial^2 \widehat{Q}}{\partial P(L_0)^2} & -\lambda & -\lambda & 0 & 1 - P(S) - P(G) \\\\
		- \lambda & \dfrac{\partial^2 \widehat{Q}}{\partial P(G)^2} & 0 & \lambda & P(R) - P(L_0)\\\\
		-\lambda & 0 & \dfrac{\partial^2 \widehat{Q}}{\partial P(S)^2} & 0 & -P(L_0)\\\\
		0 & \lambda & 0 & \dfrac{\partial^2 \widehat{Q}}{\partial P(R)^2} & -1+P(G)\\\\
		\lambda \cdot \dfrac{\partial c}{\partial P(L_0)} & \lambda \cdot \dfrac{\partial c}{\partial P(G)} & \lambda \cdot \dfrac{\partial c}{\partial P(S)} & \lambda \cdot \dfrac{\partial c}{\partial P(R)} & c,
	\end{array}\right]$}
\end{equation}
where {--}
\begin{equation}\label{eq:EM_interior:dc}
	\begin{array}{lcl}
		\dfrac{\partial c}{\partial P(L_0)} & = & 1 - P(S) - P(G),\\
		\dfrac{\partial c}{\partial P(G)} & = & P(R) - P(L_0),\\
		\dfrac{\partial c}{\partial P(S)} & = & -P(L_0),\\
		\dfrac{\partial c}{\partial P(R)} & = & - 1 + P(G).
	\end{array}
\end{equation}
Note from (\ref{eq:BKT:dQdL_0}) - (\ref{eq:BKT:dQdR}) that each first partial derivative of $\widehat{Q}$ has the following form {--}
\begin{equation}\label{eq:EM_interior:dQ}
	\dfrac{\partial \widehat{Q}}{\partial P(\boldsymbol{\cdot})} = \dfrac{A}{P(\boldsymbol{\cdot})} - \dfrac{B}{1 - P(\boldsymbol{\cdot})}
\end{equation}
with some values of $A$ and $B$ independent of $P(\boldsymbol{\cdot})$. Therefore, all corresponding second partial derivatives have the following form {--}
\begin{equation}\label{eq:EM_interior:dQ2}
	\dfrac{\partial^2 \widehat{Q}}{\partial P(\boldsymbol{\cdot})^2} = -\dfrac{A}{P(\boldsymbol{\cdot})^2} - \dfrac{B}{(1 - P(\boldsymbol{\cdot}))^2}. 
\end{equation}

To summarize, we begin with $\mu=\mu_1$ and find zero of function $F(\mu_1)$ starting with some random initial guesses $(\theta_1(\mu_1), \lambda_1(\mu_1))$ and update them using rule (\ref{eq:EM_interior:Newton_update}) and formulae (\ref{eq:EM_interior:Newton_method}), (\ref{eq:EM_interior:F}) - (\ref{eq:EM_interior:dQ2}), (\ref{eq:BKT:dQdL_0}) - (\ref{eq:BKT:dQdR}) until it converges to values $(\theta_k(\mu_1), \lambda_k(\mu_1))$ for some $k$. Then we apply the same procedure to find zero of function $F(\mu_2)$ using $(\theta_k(\mu_1), \lambda_k(\mu_1))$ as initial guesses. We continue until we converge to $(\theta_{k'}(\mu_W), \lambda_{k'}(\mu_W)) $, the solution of $F(\mu_W) = 0$, which maximizes (\ref{eq:EM_interior:B}) and is solution of (\ref{eq:EM_interior:KKT_problem}).
\section{Demonstrating the EM-Interior Point Method on Simulated Data}
\begin{figure}[htbp]
   \centering
   \includegraphics[width=0.49\textwidth]{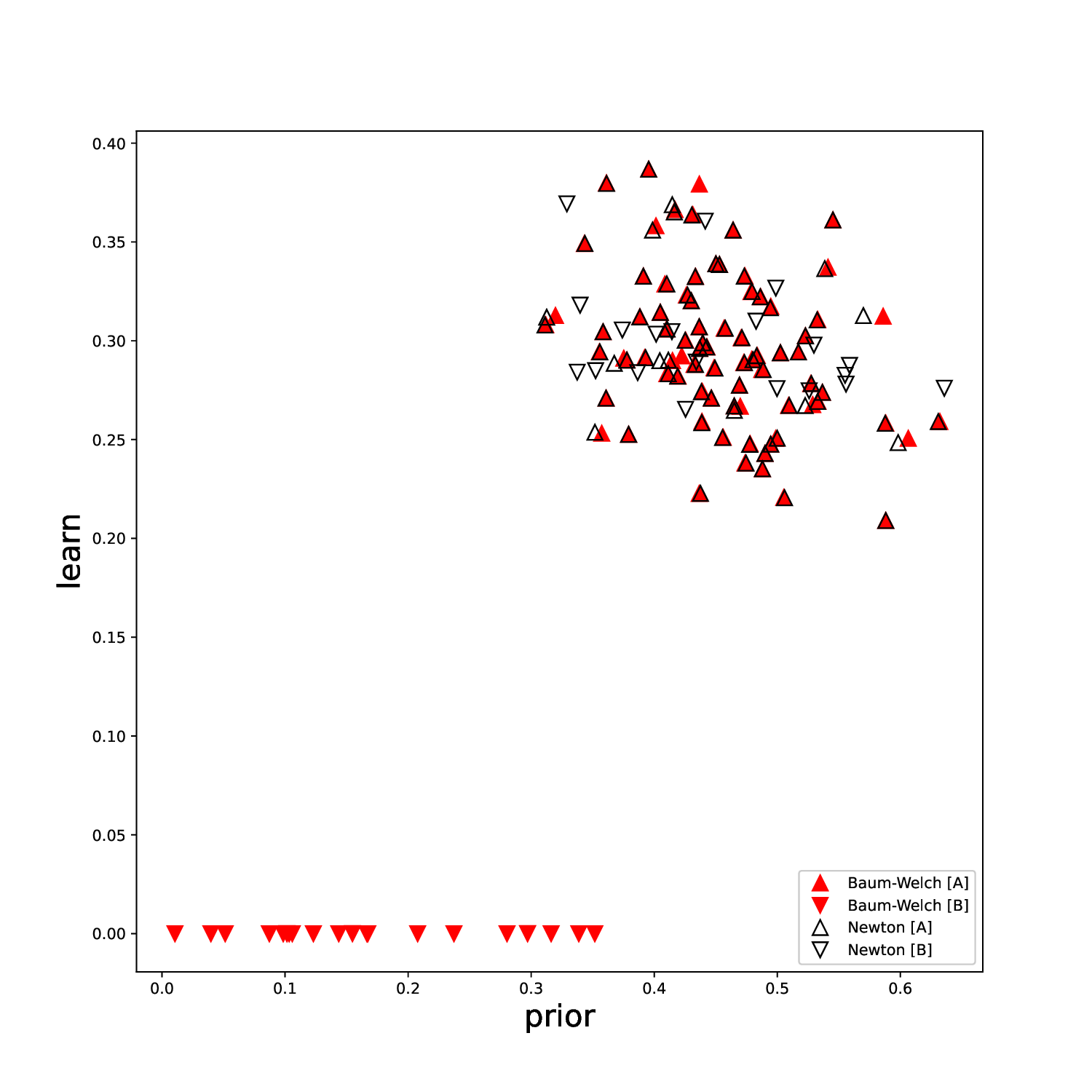} 
   \includegraphics[width=0.49\textwidth]{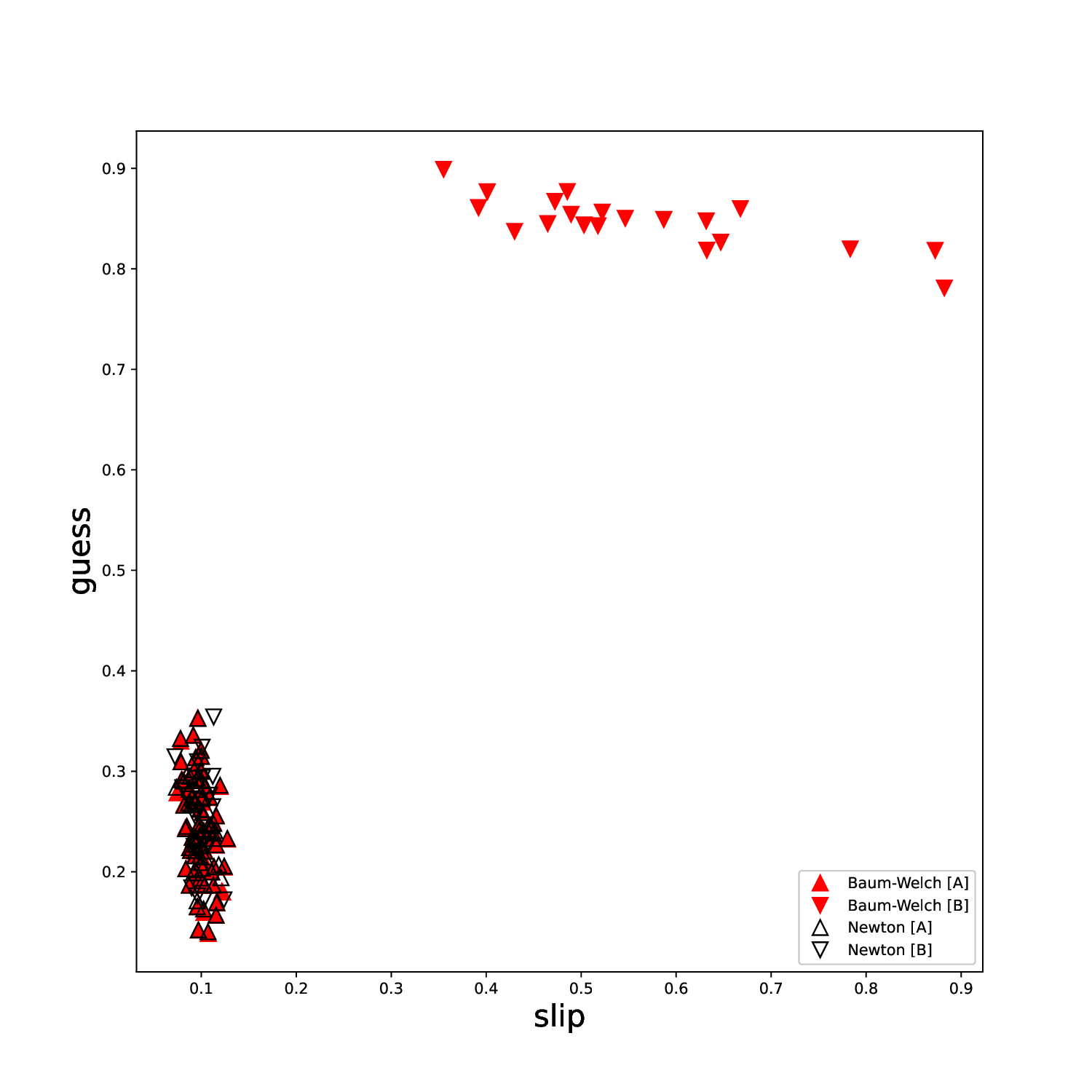} 
   \caption{BKT model fitted to 100 simulated datasets using classical Baum-Welch algorithm (red triangles) and proposed EM-Interior Point method (hollow black triangles). The data was simulated using the same value of parameters: $P(L_0)=0.45$, $P(R)=0.3$, $P(S)=0.1$, and $P(G)=0.25$. For each dataset both algorithms used the same random initial parameter guesses. There were 80 datasets for which fitted parameters satisfied the parameter conditions for both algorithms (datasets [A]; red upward triangles and hollow black upward triangles). For the remaining 20 datasets (datasets [B]), the fitted parameters did not satisfy the conditions when Baum-Welch algorithm was used (red downward triangles), but they were rescued by EM-newton method (hollow black downward triangles).}
    \label{fig:different_datasets}
\end{figure}

\begin{figure}[htbp]
   \centering
   \includegraphics[width=0.49\textwidth]{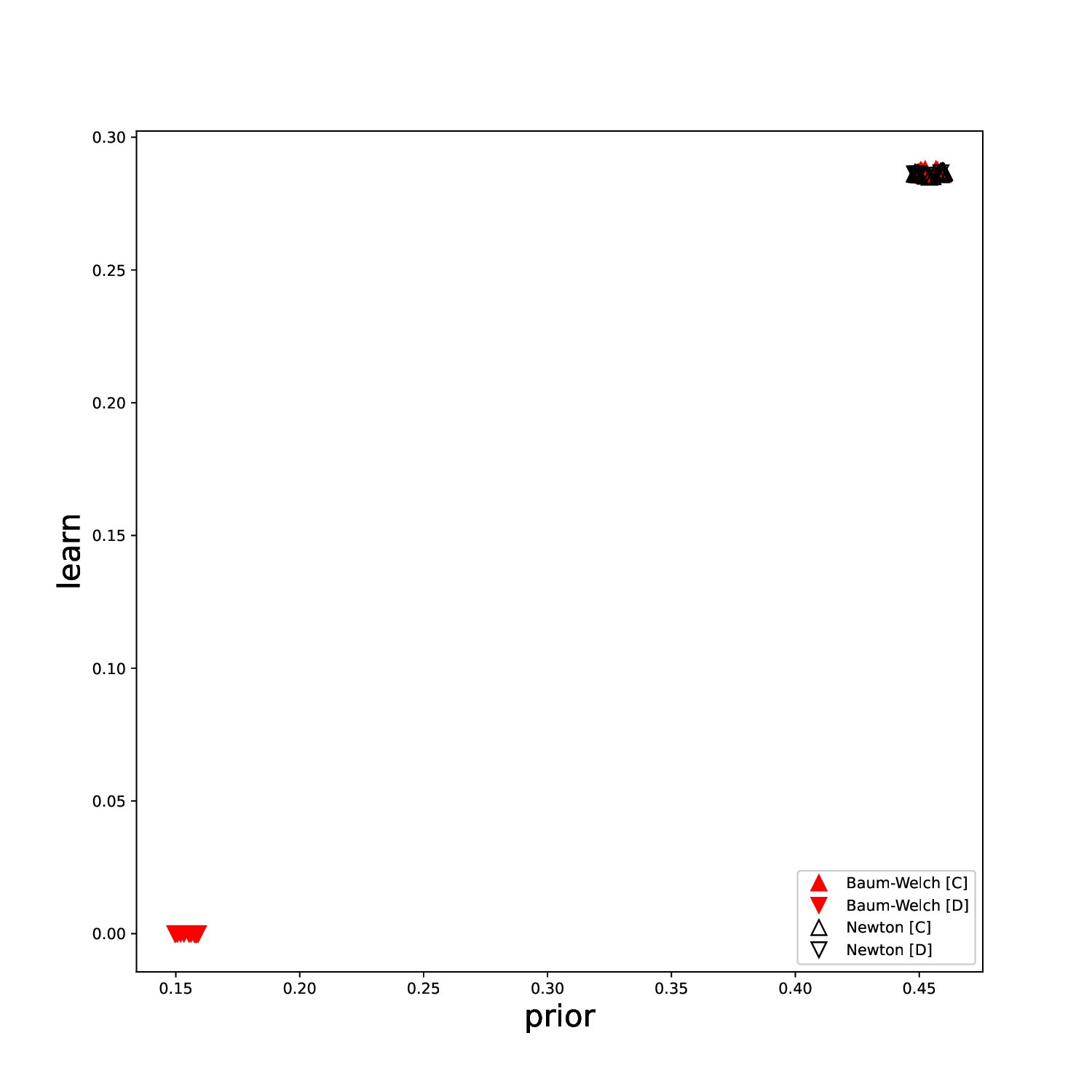}
   \includegraphics[width=0.49\textwidth]{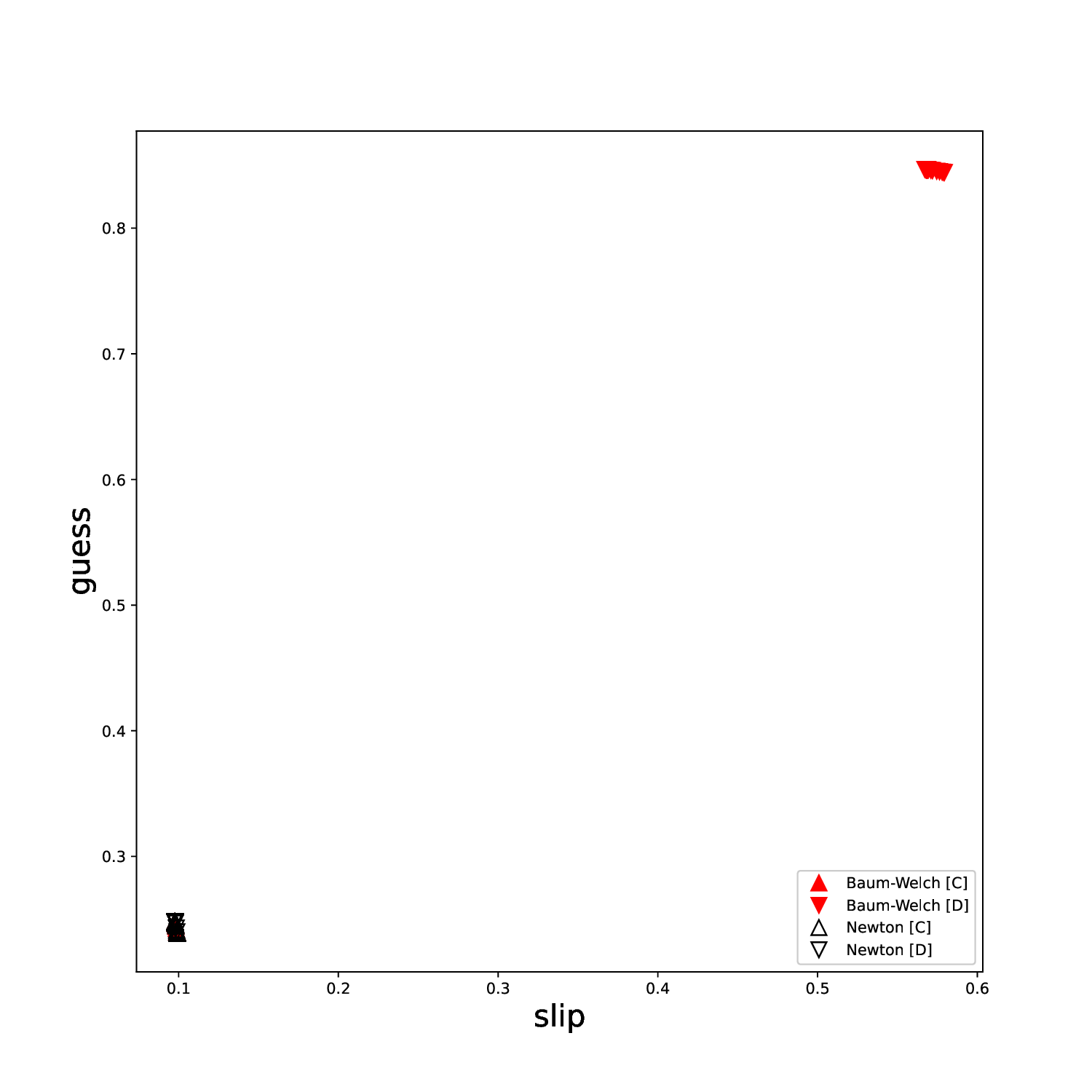} 
   \caption{BKT model fitted to the same dataset from 100 different initial parameter guesses using classical Baum-Welch algorithm (red triangles) and proposed EM-Newton method (hollow black triangles). The data was simulated using the same parameters values as in Figure \ref{fig:different_datasets}. There were 80 initial guesses that converged to parameters satisfying the conditions for both algorithms (datasets [C]; upward triangles), and 20 initial guesses that converged to parameters satisfying the conditions only for EM-Newton algorithm (datasets [D]; downward triangles). Note that for this simulation, the parameter estimates have lower accuracy but higher precision.}
    \label{fig:same_dataset}
\end{figure}

Figures \ref{fig:different_datasets} and \ref{fig:same_dataset} produce comparisons between Baum-Welch and EM-Interior Point method. The data is made available along with the paper.

\section{Discussion}
The paper first defines the list of constraints on the BKT parameter space. One question that may arise in the reader's mind is around the validity of the constraints imposed on the BKT parameters in practice. While the justifications for the constraints are mentioned in the text, they also assume that the questions are ``well-designed". It follows, therefore, that using this process, it is possible to address the complementary issue of identifying poorly performing KCs as those for whom these constraints are violated and flag them to learning designers with appropriately recommended fixes. For example, $P(R) = 1$ being true could mean that the learning experience is not connected to the KC since it is leading to proficiency regardless of mastery. While $1 - P(S) < P(G)$ could tell us the question is worded in such a way that leads to overthinking, i.e., skillful learners are less likely to answer it correctly than unskillful learners guessing the answer by chance.

Ultimate, we derived an algorithm that converges to a set of parameters that are guaranteed to meet the constrains. Additionally, we compared our algorithm to the classical Baum-Welch algorithm used to estimate parameters of Hidden Markov Models, including BKT. We demonstrated that both algorithms converge to similar values of parameters in case when these values satisfy the derived conditions. We also demonstrated that Baum-Welch algorithm occasionally converges to the values of parameters than are neither close to the true values nor satisfying of the conditions, with our algorithm being able to rescue those cases. Although a single run of the Baum-Welch algorithm is less computational heavy than a single run of our algorithm (ours requires for the Newton method to converge on each iteration), the Baum-Welch algorithm is often run multiple times with different initial conditions after post-hoc finding degenerate parameters. Our algorithm can be run once and, therefore, be less computational heavy overall.

Finally, let us also notice that the derivation approach described in the paper can be followed to devise an algorithm subject to a different set of constraints as well, so long as the set of constraints remain tractable.

\section{Conclusion and Future Work}
This paper derives constraints that can be imposed on the BKT parameter space in succinct form from first principles. Then, a new Expectation-Maximization algorithm using the Interior-Point Method is introduced that produces parameters subject to those constraints and is, therefore, guaranteed to produce valid, i.e., non-degenerate parameters. In future work, we wish to demonstrate the use of this algorithm to identify poorly performing KCs and recommend appropriate fixes to learning designers. We further intend to derive the algorithm for extensions to BKT that can still use Expectation-Maximization such as the addition of individual item difficulty \cite{pardos2011kt}, individualization \cite{pardos2010modeling, yudelson2013individualized}, time between attempts \cite{qiu2011does}, or forgetting \cite{khajah2016deep} parameters. More complex extensions such as BKT+ incorporate many of these, but, start to commensurately require more complex methods such as Markov Chain Monte Carlo (MCMC) or even deep learning \cite{gervet2020deep} which could make such first principles derivation untenable.

\newpage
\printbibliography

\newpage
\appendix
\section{Baum-Welch Algorithm}\label{apdx:Baum}
We have {--}
\begin{multline}
	P(y, x | \theta) = \prod\limits_{d=1}^D P(y^{(d)}, x^{(d)} | \theta) = \prod\limits_{d=1}^D \Biggl( (1 - P(L_0))^{(1 - x^{(d)}_1)} \cdot P(L_0)^{x^{(d)}_1}\\
	\times \prod\limits_{t=1}^{T^{(d)}} \left[ P(G)^{y^{(d)}_t} \cdot (1 - P(G))^{1 - y^{(d)}_t}) \right]^{1 - x^{(d)}_t} \cdot \left[ P(S)^{1 - y^{(d)}_t} \cdot (1 - P(S))^{y^{(d)}_t} \right]^{x^{(d)}_t}\\
	\times \left.\prod\limits_{t=1}^{T^{(d)} - 1} \left[ (1 - P(R))^{1 - x^{(d)}_{t+1}} \cdot P(R)^{x^{(d)}_{t+1}} \right]^{1 - x^{(d)}_t} \cdot \left[ x^{(d)}_{t+1} \right]^{x^{(d)}_t} \right).
\end{multline}
Thus, (\ref{eq:EM:Q_hat_discrete}) becomes {--}
\begin{multline}\label{eq:BKT:Q}
	\widehat{Q}(\theta | \theta^*) = \sum\limits_{x\in\mathcal{X}} \left[ \sum\limits_{d=1}^D \left(
	(1 - x^{(d)}_1) \cdot \log (1 - P(L_0)) + x^{(d)}_1 \cdot \log P(L_0) \right) \right. \\
	+ \sum\limits_{t = 1}^{T^{(d)}} \left( (1 - x^{(d)}_t) \cdot y^{(d)}_t \cdot \log P(G) + (1 - x^{(d)}_t) \cdot (1 - y^{(d)}_t) \cdot \log (1 - P(G)) \right) \\
	+ \sum\limits_{t=1}^{T^{(d)}} \left( x^{(d)}_t \cdot (1 - y^{(d)}_t) \cdot \log P(S) + x^{(d)}_t \cdot y^{(d)}_t \cdot \log (1 - P(S)) \right) \\
	+ \sum\limits_{t=1}^{T^{(d)} - 1} \left( (1 - x^{(d)}_t) \cdot (1 - x^{(d)}_{t+1}) \cdot \log (1 - P(R)) + (1 - x^{(d)}_t) \cdot x^{(d)}_{t+1} \cdot \log P(R) \right) \\
	\left. + \sum\limits_{t=1}^{T^{(d)} - 1} x^{(d)} \log x^{(d)}_{t+1} \right] \cdot P(y, x | \theta^*)
\end{multline}
The maximum of the function can be found by finding extremum of $\widehat{Q}(\theta | \theta^*)$. For $P(L_0)$ we have
\begin{multline}\label{eq:BKT:dQdL_0}
	\dfrac{\partial \widehat{Q}}{\partial P(L_0)} = \dfrac{\partial}{\partial P(L_0)} \sum\limits_{x \in \mathcal{X}} \sum\limits_{d=1}^D  \left( (1 - x^{(d)}_1) \cdot \log (1 - P(L_0)) + x^{(d)}_1 \cdot \log P(L_0) \right)\\
	\times P(y, x | \theta^*) = \dfrac{\partial}{\partial P(L_0)} \left( \sum\limits_{d=1}^D \log (1 - P(L_0)) \cdot P(x^{(d)}_1 = 0, y | \theta^*) \right.\\
	\left. + \sum\limits_{d=1}^D \log P(L_0) \cdot P(x^{(d)}_1 = 1, y | \theta^*) \right) \\
	= \dfrac{\sum\limits_{d=1}^D P(x^{(d)}_1 = 1, y | \theta^*)}{P(L_0)} - \dfrac{\sum\limits_{d=1}^D P(x^{(d)}_1 = 0, y | \theta^*)}{1 - P(L_0)}.
\end{multline}
For $P(G)$ we have {--}
\begin{multline}\label{eq:BKT:dQdG}
	\dfrac{\partial \widehat{Q}}{\partial P(G)} = \dfrac{\partial}{\partial P(G)} \sum\limits_{x \in \mathcal{X}} \sum\limits_{d=1}^D \sum\limits_{t=1}^{T^{(d)}} (1 - x^{(d)}_t) \cdot \left(y^{(d)}_t \cdot \log P(G) \right. \\
	\left. + (1 - y^{(d)}_t) \cdot \log (1 - P(G)) \right) \\
	= \dfrac{\sum\limits_{d=1}^D \sum\limits_{t=1}^{T^{(d)}} y^{(d)}_t \cdot P(x^{(d)}_t = 0, y | \theta^*)}{P(G)} - \dfrac{\sum\limits_{d=1}^D \sum\limits_{t=1}^{T^{(d)}} (1 - y^{(d)}_t) \cdot P(x^{(d)}_t = 0, y | \theta^*)}{1 - P(G)}.
\end{multline}
For $P(S)$ we have {--}
\begin{multline}\label{eq:BKT:dQdS}
	\dfrac{\partial \widehat{Q}}{\partial P(S)} = \dfrac{\partial}{\partial P(S)} \sum\limits_{x \in \mathcal{X}} \sum\limits_{d=1}^D \sum\limits_{t=1}^{T^{(d)}} x^{(d)}_t \cdot \left( (1 - y^{(d)}_t) \cdot \log P(S) + y^{(d)}_t \cdot \log (1 - P(S)) \right) \\
	= \dfrac{\sum\limits_{d=1}^D \sum\limits_{t=1}^{T^{(d)}} (1 - y^{(d)}_t) \cdot P(x^{(d)}_t = 1, y | \theta^*)}{P(S)} - \dfrac{\sum\limits_{d=1}^D \sum\limits_{t=1}^{T^{(d)}} y^{(d)}_t \cdot P(x^{(d)}_t = 1, y | \theta^*)}{1 - P(S)}.
\end{multline}
And for $P(R)$ we have {--}
\begin{multline}\label{eq:BKT:dQdR}
	\dfrac{\partial \widehat{Q}}{\partial P(R)} = \dfrac{\partial}{\partial P(R)} \sum\limits_{x \in \mathcal{X}} \sum\limits_{d=1}^D \sum\limits_{t=1}^{T^{(d)} - 1} (1 - x^{(d)}_t) \cdot \left( (1 - x^{(d)}_{t+1}) \cdot \log (1 - P(R)) \right. \\
	\left. + x^{(d)}_{t+1} \cdot \log P(R) \right) \\
	= \dfrac{\sum\limits_{d=1}^D \sum\limits_{t=1}^{T^{(d)} - 1} P(x^{(d)}_t = 0, x^{(d)}_{t+1} = 1, y | \theta^*)}{P(R)} - \dfrac{\sum\limits_{d=1}^D \sum\limits_{t=1}^{T^{(d)} - 1} P(x^{(d)}_t = 0, x^{(d)}_{t+1} = 0, y | \theta^*)}{1 - P(R)}.
\end{multline}
Setting partial derivatives (\ref{eq:BKT:dQdL_0}) - (\ref{eq:BKT:dQdR}) to zero, yields a closed form solution for the parameters:
\begin{equation}\label{eq:BKT:Baum-Welch_general}
	\begin{array}{lcl}
		P(L_0) & = & \dfrac{\sum\limits_{d=1}^D P(x^{(d)}_1 = 1, y | \theta^*)}{\sum\limits_{d=1}^D P(y | \theta^*)} = \dfrac{1}{D}  \sum\limits_{d=1}^D P(x^{(d)}_1 = 1 | y^{(d)}, \theta^*),\\\\
		P(G) & = & \dfrac{\sum\limits_{d=1}^D \sum\limits_{t=1}^{T^{(d)}} y^{(d)}_t \cdot P(x^{(d)}_t = 0 | y^{(d)}, \theta^*)}{\sum\limits_{d=1}^D \sum\limits_{t=1}^{T^{(d)}} P(x^{(d)}_t = 0 | y^{(d)}, \theta^*)},	\\\\
		P(S) & = & \dfrac{\sum\limits_{d=1}^D \sum\limits_{t=1}^{T^{(d)}} (1 - y^{(d)}_t) \cdot P(x^{(d)}_t = 1 | y^{(d)}, \theta^*)}{\sum\limits_{d=1}^D \sum\limits_{t=1}^{T^{(d)}} P(x^{(d)}_t = 1 | y^{(d)}, \theta^*)},\\
		P(R) & = & \dfrac{\sum\limits_{d=1}^D \sum\limits_{t=1}^{T^{(d)} - 1} P(x^{(d)}_t = 0, x^{(d)}_{t+1} = 1 | y^{(d)}, \theta^*)}{\sum\limits_{d=1}^D \sum\limits_{t=1}^{T^{(d)} - 1} P(x^{(d)}_t = 0 | y^{(d)}, \theta^*)}.
	\end{array}
\end{equation}

Let us now describe and algorithm to find probabilities in (\ref{eq:BKT:Baum-Welch_general}). The hidden proficiency process $X$ is Markov, therefore we can define a transition matrix $A=\{a_{ij}\}=\{P(X^{(d)}_t = j | X^{(d)}_{t - 1} = i)\}$ for $i=0,1$ and $j=0,1$:
$$A = \left[
\begin{array}{cc}
	1 - P(R) & P(R)\\
	0 & 1
\end{array}
\right],$$
and a so-called emission matrix $B=\{b_j(i)\}=\{P(Y^{(d)}_t = i | X^{(d)}_t = j)\}$ for $i=0,1$ and $j=0,1$:
$$B = \left[
\begin{array}{cc}
	1 - P(G) & P(G)\\
	P(S) & 1 - P(S)
\end{array}
\right].$$
And a vector of initial states $\pi = \{\pi_i\}=\{P(X^{(d)}_0=i)\}$ for $i=0,1$:
$$ \pi = \left[
\begin{array}{c}
	1 - P(L_0)\\
	P(L_0)
\end{array}\right].$$

Starting with some random initial guess for parameters $P(S)$, $P(G)$, $P(R)$, $P(L_0)$, denoted as vector $\theta$, we compute a so-called Forward Procedure for $d=1,\cdots,D$ and $m=0,1$:
\begin{equation}\label{eq:BKT:alpha_general}
	\alpha^{(d)}_i(t) = P\left(\left.Y^{(d)}_1 = y^{(d)}_1, Y^{(d)}_2 = y^{(d)}_2, \cdots, Y^{(d)}_t = y^{(d)}_t, X^{(d)}_t=i \right| \theta \right),
\end{equation}
by recursive formulae {--}
\begin{equation}\label{eq:BKT:alpha_recursion}
	\begin{array}{rcl}
		\alpha^{(d)}_i(1) & = & \pi_i \cdot b_i(y^{(d)}_1),\\\\
		\alpha^{(d)}_i(t + 1) & = & b_i(y^{(d)}_{t+1}) \cdot \left(\alpha^{(d)}_0(t) \cdot a_{0i} + \alpha^{(d)}_1(t) \cdot a_{1i}\right),
	\end{array}
\end{equation}
and a so-called Backward Procedure {--}
\begin{equation}\label{eq:BKT:beta_general}
	\beta^{(d)}_i(t) = P\left(\left.Y^{(d)}_{t+1} = y^{(d)}_{t+1}, Y^{(d)}_{t+2} = y^{(d)}_{t+2}, \cdots, Y^{(d)}_T = y^{(d)}_T \right| X^{(d)}_t=i, \theta \right),
\end{equation}
by recursive formulae {--}
\begin{equation}\label{eq:BKT:beta_recursion}
	\begin{array}{rcl}
		\beta^{(d)}_i(T) & = & 1,\\\\
		\beta^{(d)}_i(t) & = & \beta^{(d)}_0(t+1) \cdot a_{i0} \cdot b_0(y^{(d)}_{t+1}) + \beta^{(d)}_1(t+1) \cdot a_{i1} \cdot b_1(y^{(d)}_{t+1}).
	\end{array}
\end{equation}
Then we can define
\begin{multline}\label{eq:BKT:gamma}
	\gamma^{(d)}_i(t) = P\left(\left. X^{(d)}_t = i \right| y^{(d)}, \theta \right) = \dfrac{P\left(\left. X^{(d)}_t = i, y^{(d)} \right| \theta \right)}{P\left(\left. y^{(d)} \right| \theta\right)}\\
	= \dfrac{\alpha^{(d)}_i(t) \cdot \beta^{(d)}_i(t)}{\sum\limits_{k=0}^1 \alpha^{(d)}_k(t) \cdot \beta^{(d)}_k(t)},
\end{multline}
and
\begin{multline}\label{eq:BKT:xi}
	\xi^{(d)}_{ij}(t) = P\left(\left. X^{(d)}_t = i, X^{(d)}_{t+1} = j \right| y^{(d)}, \theta \right) = \dfrac{P\left(\left. X^{(d)}_t = i, X^{(d)}_{t+1} = j, y^{(d)}\right| \theta \right)}{P\left(\left. y^{(d)}\right| \theta \right)} \\
	=\dfrac{\alpha^{(d)}_i(t) \cdot a_{ij} \cdot b_j(y^{(d)}_{t+1}) \cdot \beta^{(d)}_j(t+1)}{\sum\limits_{k=0}^1 \sum\limits_{w=0}^1 \alpha^{(d)}_k(t) \cdot a_{kw} \cdot b_w(y^{(d)}_{t+1}) \cdot \beta^{(d)}_w(t+1)}.
\end{multline}

Therefore, using (\ref{eq:BKT:alpha_general}) - (\ref{eq:BKT:xi}), solution (\ref{eq:BKT:Baum-Welch_general}) becomes
\begin{equation}\label{eq:BKT:Baum-Welch}
	\begin{array}{lcl}
		P(L_0) & = & \dfrac{1}{D} \sum\limits_{d=1}^D \gamma^{(d)}_1 (1),\\\\
		P(G) & = & \dfrac{\sum\limits_{d=1}^D \sum\limits_{t=1}^{T^{(d)}} y^{(d)}_t \cdot \gamma^{(d)}_0(t)}{\sum\limits_{d=1}^D \sum\limits_{t=1}^{T^{(d)}} \gamma^{(d)}_0(t)},\\\\
		P(S) & = & \dfrac{\sum\limits_{d=1}^D \sum\limits_{t=1}^{T^{(d})} (1 - y^{(d)}) \cdot \gamma^{(d)}_1(t)}{\sum\limits_{d=1}^D \sum\limits_{t=1}^{T^{(d})} \gamma^{(d)}_1(t)},\\\\
		P(R) & = & \dfrac{\sum\limits_{d=1}^D \sum\limits_{t=1}^{T{(d)} - 1} \xi^{(d)}_{01}(t)}{\sum\limits_{d=1}^D \sum\limits_{t=1}^{T{(d)} - 1} \gamma^{(d)}_0(t)}.
	\end{array}
\end{equation}

\end{document}